\documentclass[10pt, conference, letterpaper]{IEEEtran}
\pdfoutput=1

\usepackage{cite}
\usepackage{subfigure}
\usepackage{url}
\usepackage{verbatim,epsf}
\usepackage[caption=false,font=footnotesize]{subfig}
\usepackage{enumitem}

\usepackage{tabularx}
\usepackage{balance} 

\usepackage{amssymb}
\usepackage{amsmath}
\usepackage{graphicx}
\usepackage{epstopdf}

\usepackage{epsfig, color}

\usepackage{comment} 
\usepackage{amsfonts} 
\usepackage{amsmath} 
\usepackage{rotating}

\usepackage{amssymb}
\usepackage{amsmath}
\usepackage{amsfonts}
\usepackage{wasysym}

\usepackage{algorithm}
\usepackage{algorithmic}

\usepackage{color}
 
\newsavebox{\ieeealgbox}

\newlength\myindent
\IEEEaftertitletext{\vspace{-0.28in}}

\begin{document}

\title{\huge
	{A Deep Reinforcement Learning Approach to Multi-component Job Scheduling in Edge Computing}}
\author{ 
	\IEEEauthorblockN{Zhi Cao\IEEEauthorrefmark{1}, Honggang Zhang\IEEEauthorrefmark{1},
		Yu Cao\IEEEauthorrefmark{2}, Benyuan Liu\IEEEauthorrefmark{2}\\ 
		\IEEEauthorrefmark{1} UMass Boston, Boston, MA. Email: \{zhi.cao001,honggang.zhang\}@umb.edu \\
		\IEEEauthorrefmark{2} UMass Lowell, Lowell, MA. Email: \{yu\_cao@uml.edu, bliu@cs.uml.edu\}
	}
}

\maketitle
\pagestyle{plain}

\begin{abstract}
We are interested in the optimal scheduling of a collection of multi-component application jobs in an 
edge computing system that consists of geo-distributed edge computing nodes connected through 
a wide area network. The scheduling and placement of application jobs in an edge system is challenging 
due to the interdependence of multiple components of each job, and the communication delays 
between the geographically distributed data sources and edge nodes and their dynamic availability.
In this paper we explore the feasibility of applying Deep Reinforcement Learning (DRL) based design 
to address these challenges. We introduce a DRL actor-critic algorithm that aims to find an optimal 
scheduling policy to minimize average job slowdown in the edge system. 
We have demonstrated through simulations that our design outperforms a few existing algorithms, 
based on both synthetic data and a Google cloud data trace. 
\end{abstract}

\section{Introduction}

Edge computing is an emerging computing paradigm that can reduce network transmission cost of the large amount 
of data generated by edge devices (e.g., IoT sensors, mobile devices) and provide low-latency service to 
end users, through processing computing jobs at the edge and close to end users \cite{fog17,lavea17,cha17,multiaccess17}.
Edge computing is well suited for the bandwidth hungry, latency-sensitive, 
and location-aware applications, e.g., real-time monitoring, cognitive assistance, and self-driving vehicles.

An important issue in edge computing is the allocation of 
resources to the competing demands from different applications 
\cite{jobdispatch17,zc182,zc181, joint18, vr18, sdn18, vnf18, heteroedge19}. 
Existing studies usually formulate edge resource allocation as optimization problems 
and design heuristics to solve the problems due to their hardness. 
Usually the proposed heuristics are only applicable for some particular types of edge computing environments.
It is difficult to develop accurate models to solve resource allocation and job scheduling problems 
in edge systems due to the complexity of those systems.

Typically an edge application job consists of multiple task components that need to be processed 
at different edge computing nodes that are geo-distributed \cite{zc182, bahreini2017efficient}. 
For example, an environment scientist may need to compare 
the environmental statistics of several geo-locations. A smart city's traffic control system needs to
compare real-time statistics on the automobiles in a number of places in a city, 
and in this case a video analysis component running at an edge node 
needs to process the data collected on-site and 
then sends its result back to another edge node or the cloud for further analysis.  

In this paper, we explore the feasibility of applying Deep Reinforcement Learning \cite{large15} techniques to 
the scheduling/placement of a collection of application jobs, 
each consisting of multiple-components to be executed in different edge nodes that are geo-distributed far apart.
We model the scheduling problem as a Markov Decision Process (MDP) \cite{rl}, 
and introduce a Deep Reinforcement Learning (DRL) algorithm to find a good scheduling policy 
in an edge computing system.  
DRL has been applied recently in distributed systems and computer networks \cite{maorm16, maona17,li2018model, deepconf18,multiagent19}. 
Instead of formulating problems with accurate mathematical models, 
DRL approximates complex stochastic dynamic systems as Deep Neural 
Networks (DNN) and makes optimal system scheduling from its experience. 
In DRL literature, there are primarily two kinds of widely used policies: value-based and actor-based\cite{large15}. 
Our design leverages 
an actor-based method,
that uses Deep Deterministic Policy Gradient\cite{continuous15} method 
to obtain the advantages of both value-based methods and actor-based methods.



Our main contributions are as follows:
\begin{enumerate}
	\item We model the scheduling of a collection of multi-component application jobs 
	in an edge computing system as a MDP problem, and introduce 
	a Deep Reinforcement Learning (DRL) model to solve the problem. Our DRL model 
	captures the status of edge nodes' resource allocation to various components of different jobs, and the status
	of data transmissions between edge nodes. Our DRL policy training algorithm 
	utilizes the recent DRL actor-critic method to find a near optimal online scheduling policy. 

	
	\item Through simulation of synthetic data and a real-world Google cloud data trace, 
	we show that our proposed DRL design outperforms several existing algorithms 
	including Shortest Job First (SJF), Least Bytes First (LBF), and random scheduling algorithm. 
	
\end{enumerate}

Our paper is organized as follows. Section \ref{sec_related} discusses related work. 
Section \ref{sec_model} presents our system design based on Deep Reinforcement Learning. 
Section \ref{sec_eval} presents our evaluation results. The paper concludes at
Section \ref{sec_concl}.

\section{Related Work}\label{sec_related}

Resource allocation and job scheduling is a fundamental problem in edge computing 
\cite{jobdispatch17, joint18, hard18, zc181, zc182,vr18,sdn18,container2019, eco2018,zc182,bahreini2017efficient,vnf18}.
Existing work has shown the importance and challenges of designing efficient application job assignment
and scheduling as many edge applications need coordinated computation 
and communication between multiple components (e.g., containers, micro-services)
assigned and scheduled on geo-distributed edge nodes. 

In addition, recently Deep Reinforcement Learning has been applied in studying resource allocation problems
in distributed systems.
Mao et al. \cite{maorm16, maona17} leverage DRL techniques to solve a multi-resource cluster scheduling problem,
and they find that their RL agents sometimes can perform better than heuristics.
\cite{dp17} presents an adaptive method that uses a sequence-to-sequence model to optimize device placements. 
\cite{li2018model} investigates a model-free DRL 
to minimize the average end-to-end tuple processing time.
\cite{deepconf18} develops a DRL-based framework, DeepConf, to automatically learn and implement 
a wide range of data center networking techniques that replace existing heuristic-based approaches.
\cite{multiagent19} studies the problem of cooperative decision making from a collection of agents. 
\cite{mao2019} studies the stream processing in a cloud data center, but our work studies the edge computing applications 
for which there are communication delays between different edge computing nodes 
that work on the different components of a job. 
In addition, our system design is not restricted to jobs with DAG structures \cite{mao2019}. 

\section{System Model}\label{sec_model}

\subsection{Preliminaries}

We consider an edge computing system that consists of edge nodes 
that are geo-distributed and connected through a communication network, 
and a cloud data center that can conduct aggregate data analytics  
and help with the communication between edge nodes. 
Let $\mathcal{V}=\{v_1, v_2, ..., v_{|\mathcal{V}|}\}$ denote the set of edge nodes.
Each edge node is a cluster of computing servers that are co-located with 
a base station or in a physical location specified by the edge computing system (e.g., 
a community center, a library). A data communication network is formed between the edge nodes
of the system. 

The edge system serves the application jobs submitted by many clients.
Each job has multiple task components. Different task components of a job may be
executed by the system at different locations and at different times.
There is an execution order among task components of a job, i.e., 
a component that follows some other component of the same job cannot start unless the other component
has finished execution. 

Assume that a job scheduler resides on the cloud part of the edge system.
As soon as a job arrives at the system, its meta data (such as its number of components
and where they are to be executed) gets submitted to the scheduler. 
The scheduler then 
uses a placement and scheduling policy to schedule the various components of submitted
application jobs on different edge nodes to optimize some system performance metrics. 

We discretize the system into $T$ time slots or steps with $t=1, ..., T$.
Let $\mathcal{G}$ denote the set of application jobs.
Consider a job $g$ (i.e., $g\in \mathcal{G}$).
Let $\mathbf{g} = \{g_1, ..., g_{|\mathbf{g}|}\}$ denote 
the set of all task components of job $g$, for which component $g_i$ should be executed before $g_{i+1}$.
Let $t_{g,arrv}$ denote the time slot in which job $g$ arrives at the system.


A task component is executed at an edge node that is close to its data input (e.g., 
from real-time video surveillance systems, or IoT sensors in a smart city). 
Transferring a task component to the cloud for execution 
is not an option due to the high network transmission cost. 
Let $\mathbf{v}_{g_j}$ denote the set of edge nodes that 
are able to execute task component $g_j$ of application job $g$, 
with $\mathbf{v}_{g_j} \subseteq \mathcal{V}$.

\begin{figure}[!h]
	\centering
	\includegraphics[width=2.5in]{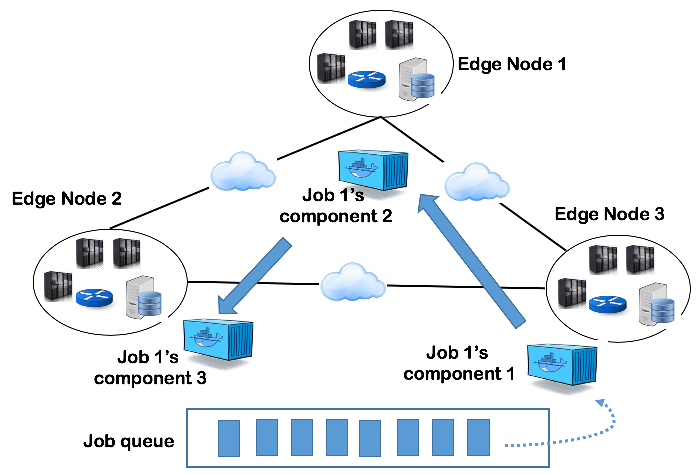}
	\caption{Example of the interdependence and order of a job's components. 
	The job is also shown as the blue-colored job in Figure \ref{state_rep}.}
	\label{job_example}
\end{figure}

A characteristic of the system we study is the interdependence and 
execution ordering between an application job's 
multiple task components. Once a job $g$'s component $g_i$ 
finishes its execution at an edge node $u$, 
it might send data/result 
to the next component of the same job, denoted by $g_{i+1}$ at
another edge node $v$.
The delay of transmitting the data can be denoted as 
$\delta_{g_{i} g_{i+1}} = h_{g_{i} g_{i+1}} / b_{u v}$,
where $h_{g_{i} g_{i+1}}$ and $b_{u v}$ represent respectively
the size (in bytes) of the transmitted data 
and the bandwidth of the link between the two edge nodes 
$u$ and $v$
(the link can be a physical link or an overlay link). 
Figure \ref{job_example} shows an example application job that consists of 
three components.

In addition, let $c_v$ denote the computation capacity of edge node $v$. 
This capacity can be understood 
as the number of computing servers at the edge node, 
or the number of parallel executing cores at the node.  
We require that at each time slot $t$, the total consumed computation resource 
of all task components executed concurrently at 
edge node $v$ should not exceed the computation capacity $c_{v}$.

\subsection{System Optimization Formulation}


The objective of the edge system is to assign task components of all application jobs to 
edge nodes and use some scheduling policy to schedule the 
task components on all edge nodes collectively in order to minimize the total or 
average application slowdown. 
In this paper, we consider a central scheduler to schedule all application jobs
submitted to the edge system, and leave the design of distributed schedulers as 
future work. 

Let $v_{g_i}$ represent the edge node that executes task component $g_i$ of job $g$, 
and $v_{g_i} \in \mathbf{v}_{g_i}$.
Let $x_{g_{i} v_{g_i}}$ denote the assignment of component $g_i$ 
on an edge node $v_{g_i}$.
Here $x_{g_{i} v_{g_i}}=1$ means that $g_{i}$ is assigned to $v_{g_i}$; $x_{g_{i} v_{g_i}}=0$
otherwise.
And $t_{g_{i} v_{g_{i}}}$
denotes the scheduled starting time of component $g_i$ on node $v_{g_{i}}$.

Let $d_{g_{i} v_{g_i}}$ denote time duration needed for edge node $v_{g_i}$ to work on 
task component $g_i$. If component $g_i$ is scheduled to be executed on edge node $v_{g_i}$,
the time needed to finish $g_i$ is given by 
$t_{g_{i} v_{g_{i} }} + d_{g_{i} v_{g_{i}}  } $.

The total time that a job $g$ stays in the system is given by 
\begin{equation}
T_g = t_{g_{|\mathbf{g}|} v_{g_{|\mathbf{g}|} }} + d_{g_{|\mathbf{g}|} v_{g_{|\mathbf{g}|}}  }
- t_{g,arrv}
\end{equation}
with $t_{g_{i} v_{g_{i}}}\ge t_{g_{i-1} v_{g_{i-1}}} + d_{g_{i-1} v_{g_{i-1}}} + \delta_{g_{i-1} g_{i}}, i=|\mathbf{g}|,|\mathbf{g}|-1,...,2$ 
and $t_{g_1 v_{g_1}} \ge t_{g, arrv}$. 
Let $T_{g, base}$ represents the ideal execution time duration of job $g$ without any delay
	due to the blocking from other jobs,  then the slowdown of job $g$ is defined as 
$\phi_g = T_{g} / T_{g, base}$.

The edge system's scheduler attempts to find a placement policy $x_{g_{i} v_{g_i}}$
for the components from all jobs, and in the mean time a scheduling policy 
$t_{g_{i} v_{g_i}}$, where $g_i \in \mathbf{g}$, $g\in \mathcal{G}$
and $v_{g_i} \in \mathcal{V}$, 
in order to minimize all jobs' aggregated slowdown.
That is, the system attempts to solve the following optimization problem.
\begin{equation}
\min_{ \{t_{g_i v_{g_i}}\},\hspace{4pt} \{x_{g_i v_{g_i}}\} } 
\sum_{g=1}^{|\mathcal{G}|} \phi_g \label{platform_min} 
\end{equation}
subject to the following constraints: (1) All components of a job should follow their 
execution ordering. That is, a component's start time should be no earlier than the 
completion time of a previous component plus the data transmission delay from the previous 
component to this component, i.e., 
$t_{g_{i+1} v_{g_{i+1}}}\ge t_{g_{i} v_{g_{i}}} + d_{g_{i} v_{g_{i}}  } + \delta_{g_{i} g_{i+1}}$.
(2) A task component can only be placed on one edge node and the edge node 
must be one of the multiple edge nodes that are eligible to execute the component 
due to edge computing requirement.
(3) The resource requirements of all components (which 
might be from different jobs) that are scheduled to be executed concurrently at 
an edge node should not exceed the capacity of the edge node. Due to space limitations, 
a formal and complete optimization formulation is not presented here. 

In this paper, we focus on the scheduling of task components of all jobs at 
edge nodes and assume that the placement of the components at edge nodes are known and fixed. 
Thus for presentation clarity, we will omit edge node subscripts in the remaining of the paper
when applicable (e.g., using $d_{g_i}$ instead of $d_{g_iv_{g_i}}$).

\subsection{Dynamic Scheduling via Deep Reinforcement Learning}

\begin{figure}[!h]
	\centering
	\includegraphics[width=3.5in]{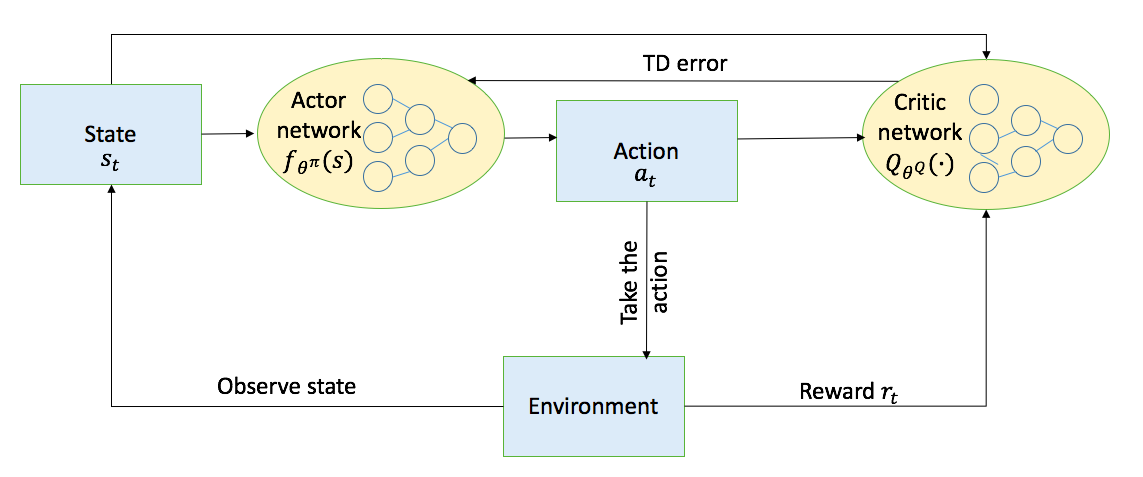}
	\caption{DRL scheduling policy.}
	\label{drl_arch}
\end{figure}

In practice, application jobs arrive at the system dynamically,
usually following some stochastic process. 
We model the online job scheduling problem (\ref{platform_min}) as a Markov Decision Problem (MDP).
To solve the problem, 
we introduce a scheduling policy based on Deep Reinforcement Learning (DRL)'s
\textit{actor-critic method} \cite{continuous15, large15}, 
where deep neural networks are used as function approximators to estimate values 
or compute a policy distribution, and the architecture of the method is illustrated in Figure \ref{drl_arch}. 
Function approximator $f_{\theta^{\pi}}(s) = a$ maps a state from state space
to action space.
An actor deep neural network computes the function $f_{\theta^{\pi}}$ and provides 
an action for a given state. 
Then a critic network $Q(s, a, \theta^{Q})$ 
returns $Q$ value for each action using the Bellman equation, parameterized by 
$\theta^{Q}$. The actor network is trained by the deterministic policy gradient method, 
with the help of the critic network. 

We next describe our DRL-based model of the edge system's state space, action space,
rewards, and our proposed training algorithm that generates an online multi-component
job scheduling policy, based on a large amount of training samples accumulated from past experience.

\bigskip
\noindent \textbf{State Space.} 
A state of the edge system in a time slot $t$ includes the current assignment of job components to edge nodes and the
resource profiles of jobs in a job queue, and the current data transmission status 
of each edge node to other edge nodes. Let $s_t$ denote the state of the system at time $t$, 
and then $s_t=(s_{am,t}, s_{trans,t}, s_{wq,t}, s_{bl,t})$. 
Note that $s_{am,t}$ denotes the assignment of the components (of various jobs) in different time slots.
That is, it includes the scheduling information of each job $g$ that arrives at time $t$ or earlier 
and for which the scheduler is able to schedule its components 
to start executing at certain time slots at different edge nodes.
The state's second element $s_{trans,t}$ denotes the bandwidth usage status of 
the transmission links of each edge node to other edge nodes (which are used to transfer data between 
neighboring task components of the same job).  
And $s_{wq,t}$ denotes the profiles of the jobs in a waiting queue. 
Furthermore $s_{bl,t}$ denotes the number of jobs in a backlog queue. 
Note that $s_{wq,t}$ encodes the interdependence and execution order of the components of each job.  
In order to have a fixed representation of system state, we choose a finite job queue with a size of $K$, and 
leave all jobs that have arrived but have not been assigned to any edge node 
and are also not able to enter the queue in the backlog.

We use images to encode the state of the edge system (similar to \cite{maorm16}). 
Specifically, $s_t$ consists of four different types of images: edge node images for $s_{am,t}$, 
network communication images for  $s_{trans,t}$, 
waiting queue images for $s_{wq,t}$, and a backlog image for $s_{bl,t}$. 
\begin{itemize}
	\item \textit{Edge node images} show the assignment of task components on edge nodes. 
	The width of an edge node image represents the amount of computation resource per time slot of that node, denoted by $\widetilde{H}$. 
	We set it to be at least as large as the maximum resource requirement per time slot of a task component in our experiments. 
	The height of an edge node image represents how far in future we can reserve an edge node's resources. 
	We set an image's height to be at least as large as the maximum time requirement (measured in
	time slots) of a task component. We use $\widetilde{T}$ to denote the height of an image. 
	
	\item \textit{Communication images} show the status of data transfers 
	between different components (of a job)
	that are executed at different edge nodes. Similar to edge node images, the width of a communication image
	(corresponding to a data transmission link between two edge nodes)
	represents the bandwidth of a link.  It is sufficient to set the height of a communication image to be 
	the amount of time needed to transmit the largest transmission data in our experiments. 

	\item \textit{Waiting queue images} encode the profiles of the jobs in a queue waiting to be scheduled. 
	Note that a job's profile includes its component set, which has the following information of each component:
	(1) its computation resource requirement per time slot; (2) the number of time slots needed for each component;
	(3) the edge node to which a component is assigned; (4) the execution order of all components of a job. 
	
	\item \textit{Backlog image} records the number of jobs that are in the backlog of the system.
\end{itemize}



\begin{figure}[!h]
	\centering
	\includegraphics[width=3.7in]{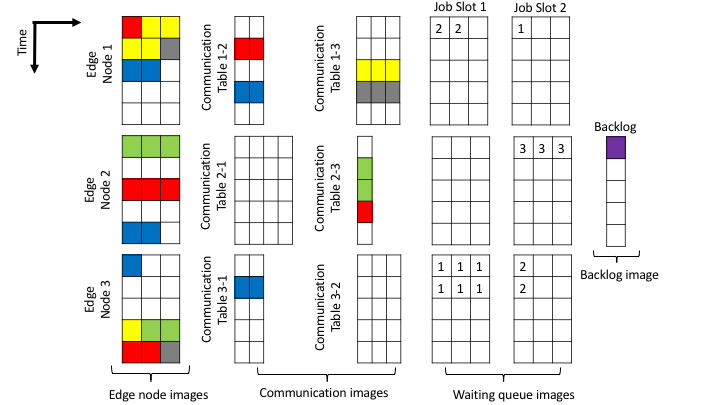}
	\caption{State representation example. }
	\label{state_rep}
\end{figure}

We now illustrate our state representation via an example shown in 
Figure \ref{state_rep}. 
In this example, there are 4 scheduled jobs with their components assigned on 
three edge nodes. There are two jobs in a waiting queue, and one job in a backlog queue.
Different jobs are distinguished with different colors. 
There are three edge node images (the leftmost three images in the figure). 
The six images in the middle represent the data communication between these edge nodes. 
These images show the bandwidth between two edge nodes and the communication status of each task component.
The width of each network image represents the capacity of a communication link 
between two edge nodes. 
We assume that at each time step, one edge node can only transfer one task component's data
to another edge node.
The top rows of the edge node images and the communication images 
correspond to time slot $0$, i.e., the current time of the system.
The rightmost images show the profiles of the jobs in the waiting queue and the number of jobs in the backlog.

In Figure \ref{state_rep}, the four jobs that have been scheduled for execution 
at three edge nodes in the next $\widetilde{T}$ time steps are marked as red, yellow, blue, and gray. 
Take the blue-colored job as an example. 
The job has three task components, and the first 
is executed in the current time slot by edge node 3 using one resource slot, 
and edge node 3 will transfer the data of this first task component to edge 
node 1 in time slot 1. Then, the second task component will be executed in time slot 2 
by edge node 1 using two resource slots. Later, edge node 2 will receive data which is transferred 
by edge node 1 in time slot 3, and will execute the third component in time slot 4 using two resource slots. 

Note that a component of a job might not be able to transfer its data to the next component
at some other edge node even if it is ready, because
of the blocking by some other job. For example, 
the gray-colored job's first task component is executed in time slot 1, but after that, 
it cannot transfer its data from edge node 1 to 
edge node 3 (which will execute the second component of the same job) immediately, 
since edge node 1 is already scheduled to transfer the data of the first component of the yellow-colored job 
in time slot 2. Then, the data of the first task component of the gray-colored job will be transferred in time slot 3. 

Recall that there is an execution ordering between components of a job. 
In the example shown in Figure \ref{state_rep}, the numbers shown in the images of 
the waiting queue represent the ordering of the task components of a job. 
For example, the job in Job Slot 1 has two task components. 
The six digits of $1$ in the bottom image of job slot 1 indicate that this image is 
the job's first component, and the two digits of $2$ in the top image of Job Slot 1 indicate
that this image is the job's second component. 

\bigskip
\noindent \textbf{Action Space.}
In each time slot $t$, the system has $K_t$ 
jobs to schedule, which results in an action space of size $2^{K_t}$. 
In order to reduce the large action space, we utilize a method similar to \cite{maorm16,li2018model}. 
At each time $t$, the scheduler's action $a_t$ consists of multiple sub-actions $a_{t,i}$. 
Each sub-action is to choose a job
in the queue $\mathcal{K}_t=\{0, 1, ..., K_t\}$. If $a_{t,i} = k$, 
the scheduler will try to schedule the $k$-th job in the queue.
We freeze the time until an invalid sub-action is found by the scheduler. 
An invalid sub-action can be: 
(1) $a_{t,i} = 0$, i.e., the scheduler does not choose any job; (2) $a_{t,i} = k$, but the $k$-th job
in the queue cannot be scheduled at $t$ due to the current status of edge nodes and/or their communication links,
or the execution ordering requirement of the components of the $k$-th job.
If the scheduler finds an invalid sub-action, it will collect all of its valid sub-actions to form action $a_t$ and the system time will move forward again.

\bigskip
\noindent \textbf{Rewards.}
Recall that the system's objective is to minimize the average 
slowdown of all jobs. 
In each time slot $t$, we let reward $r_t=\sum_{g \in \mathcal{G}_{t,remain}} - 1/T_{g, base}$, 
and $\mathcal{G}_{t, remain}$ contains at time $t$ the jobs in the waiting queue, the jobs in the 
backlog, and the jobs that have been scheduled but not completed yet. 


\bigskip
\noindent \textbf{Training Algorithm to Find Scheduling Policy}.
Our  algorithm utilizes a critic network and an actor network. 
The critic network is used only to train a actor network, not to provide 
actions to schedule jobs in the waiting queue. 
To train the actor network and the critic network in a stable 
and robust way, we use experience replay buffer $R$, as suggested in \cite{continuous15}. 
At each time step, we store a quadruple $(s_t, a_t, r_t, s_{t+1})$ as a sample in the replay buffer $R$, which is 
finite with a fixed size $Z$. 
When the replay buffer is full, the oldest 
sample will be replaced. When the networks begin to be trained, a mini-batch of 
samples from the replay buffer is chosen to update the networks. 

For each quadruple sample in the mini-batch, we first obtain an action 
$a_{t+1}'$ for the next state 
$s_{t+1}$, and obtain the estimated Q-value for this $(s_t, a_{t+1}')$ pair. As the critic  
network is also used in calculating the estimated Q-value and the actor network is also used to 
obtain $a_{t+1}$, the networks are likely to be unstable and difficult to converge. 
Thus we use target networks $f_{\theta^{\pi}}'$ and $\theta^{Q'}$ with 
soft updates to robustly 
train these two networks. 
That is, we use a target actor network to obtain $a_{t+1}'$ for next state, and 
calculate estimated Q-value $Q(s_{t+1}, a_{t+1}')'$ by a target critic network. 
Then  we set $y_t$ 
as the sum of immediate reward $r_t$ and the discounted Q-value $Q(s_{t+1}, a_{t+1}')'$ as follows:
$y_t = r_t + \gamma Q(s_{t+1}, a_{t+1}')'$.
In our experiments, $\gamma = 0.99$. 
Then the critic network is updated by minimizing a common loss,
\begin{eqnarray}
L(\theta^{Q}) = \frac{1}{M}\sum_{i=1}^{M}(y_t - Q_{s_t, a_t})^2.
\end{eqnarray}
The actor network is updated using sampled policy gradient,
\begin{eqnarray}
\bigtriangledown_{\theta^{\pi}} f \approx \frac{1}{M} \sum_{i=1}^{M}\bigtriangledown_a Q(s, a|\theta^{Q})|_{s=s_i, a=f(s_i)} \bigtriangledown_{\theta^{\pi}} f(s)|s_i.
\end{eqnarray}
The target actor network and the target critic network are update softly, 
\begin{eqnarray}
\theta^{Q'} \leftarrow \theta^{Q} + (1-\tau)\theta^{Q'}, \\
\theta^{\pi'} \leftarrow \theta^{\pi} + (1-\tau) \theta^{\pi'}.
\end{eqnarray}

\section{Evaluation}\label{sec_eval}

We next present some of our simulation results to compare our DRL scheduling policy's performance 
with some other scheduling policies.

\subsection{Experiment Setup}

We first generate two synthetic datasets to evaluate the performance of our DRL scheduling policy. 
Dataset $1$ has only two types of jobs, long job and short job. Each task component of a long job
has a time requirement of 9 time slots and a computation resource requirement of 9 slots
per time slot; for a short job's task component, these two resource requirements are 8 and 18 respectively.
For dataset 2, we randomize the time requirement and resource requirement per time slot of both long and 
short jobs' task components, but we make sure the long jobs have longer time requirements 
(i.e., larger $T_{g,base}$) than the short jobs. 
The edge node images used by DRL agents to run the simulation of dataset 1 and dataset 2 have 
$\widetilde{H} = 20$ resource slots and $\widetilde{T} = 100$ time slots. 


We also run simulation with a dataset from Google cloud trace \cite{googleclouddata}. 
We derive a task component's time duration by its \textit{SCHEDULE} flag and its \textit{FINISH} flag. 
A component's resource requirement per time slot is derived based on 
the capacity of CPU and memory that the component uses during its execution. 
We set the edge node images in the simulation of Google dataset to have 
$\widetilde{H} = 100$ resource slots per time slot and $\widetilde{T} = 200$.

We evaluate our DRL scheduling policy based on different levels of average edge node workload $\bar{L}$.
The workload of an edge node includes both computation workload and network transmission workload. 
The computation workload is calculated as the ratio of the total resources needed by all jobs in an episode
over the total available resources of the edge nodes aggregated over all time slots in an episode.
The network workload is calculated similarly. 
For the simulations of datasets 1 and 2, we let $\bar{L}$ varies from $30\%$ to $56\%$ based on job arrival rates 
$0.5,0.7$ and $0.9$.  
In our simulation based on Google dataset, 
we let $\bar{L}$ varies from $12\%$ to  $16\%$ based on job arrival rates $0.6$ and $0.8$.

We are interested in two performance metrics: 
average job completion time 
$\bar{T}_g$ (i.e., $\bar{T}_g = \frac{1}{|\mathcal{G}|} \sum_{g = 1}^{|\mathcal{G}|} T_g$), 
and average job slowdown 
$\bar{\phi}_g = \frac{1}{|\mathcal{G}|} \sum_{g=1}^{|\mathcal{G}|} T_g / T_{g, base}$.
We compare our DRL scheduling policy with three other algorithms:
Least Bytes First (LBF), Shortest Job First (SJF) and random scheduling. 
LBF gives the highest scheduling priority to the job with the least total bytes.
The total bytes of a job is calcuated as the sum of all of its components' computation sizes and
all of its transmitted data sizes. Note that the computation size of a component is the product
of its consumed number of resource slots per time slot and the number of time slots it needs. 
SJF gives the highest priority to the job with the least $T_{g, base}$.
Recall $T_{g, base}$ of a job includes only the total number of time slots needed by the job's all components 
and the transmission delays of the job. 




\subsection{Results}

\begin{figure}[!h]
	\centering
	\setlength{\epsfxsize}{2.5in}
	\epsffile{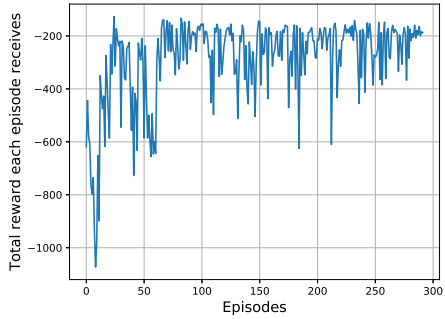}\\
	\caption{An example training process. }
	\label{converge}
\end{figure}

Figure \ref{converge} shows an example of the convergence of our DRL scheduling policy during training.
In the beginning, the total reward that each episode receives grows fast. Then, near episode 117, the curve begins 
to converge. 

\begin{figure}[htb!]
	\centerline{
		\begin{minipage}{1.7in}
			\begin{center}
				\setlength{\epsfxsize}{1.7in}
				\epsffile{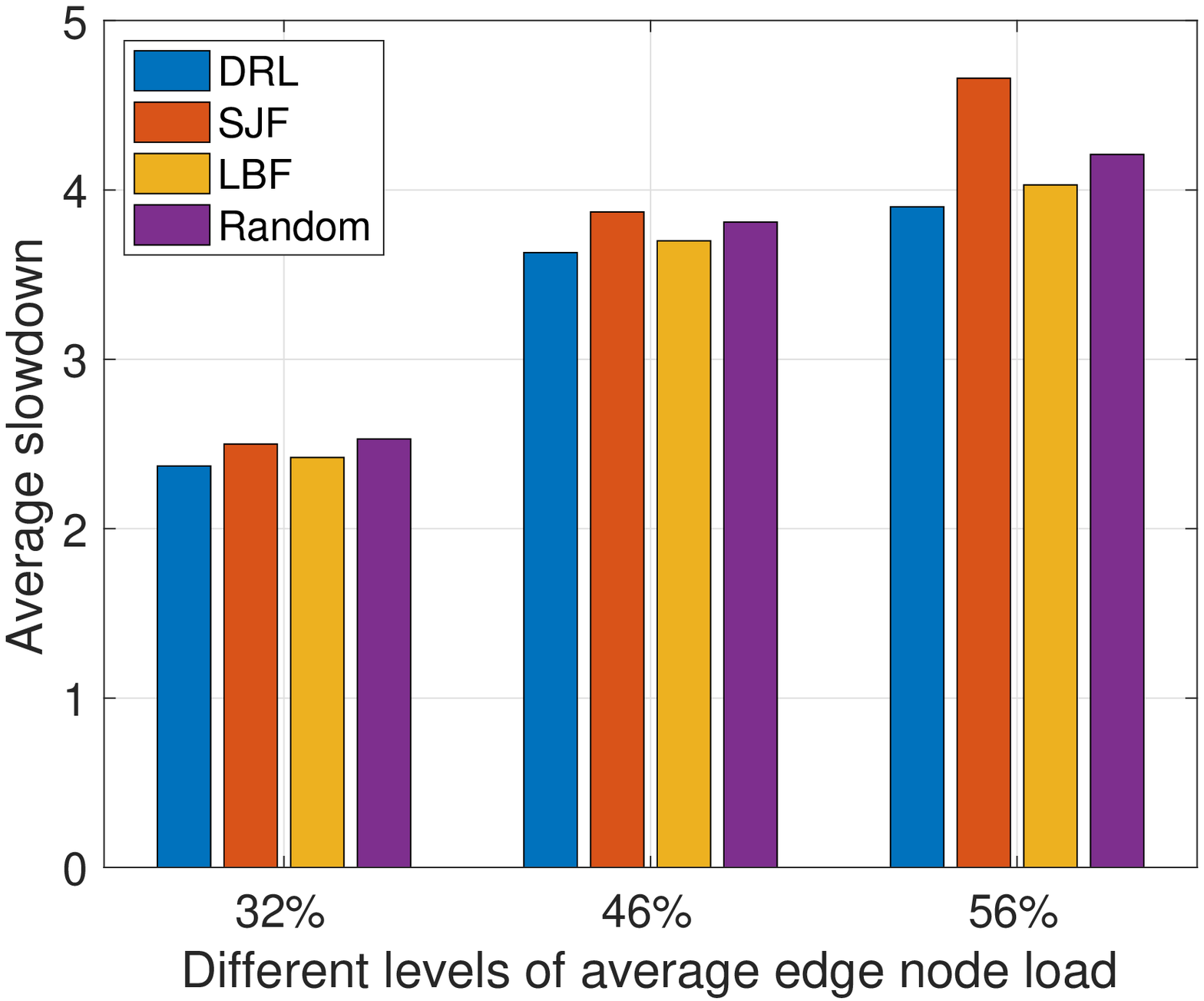}\\
				{}
			\end{center}
		\end{minipage}
		\begin{minipage}{1.7in}
			\begin{center}
				\setlength{\epsfxsize}{1.7in}
				\epsffile{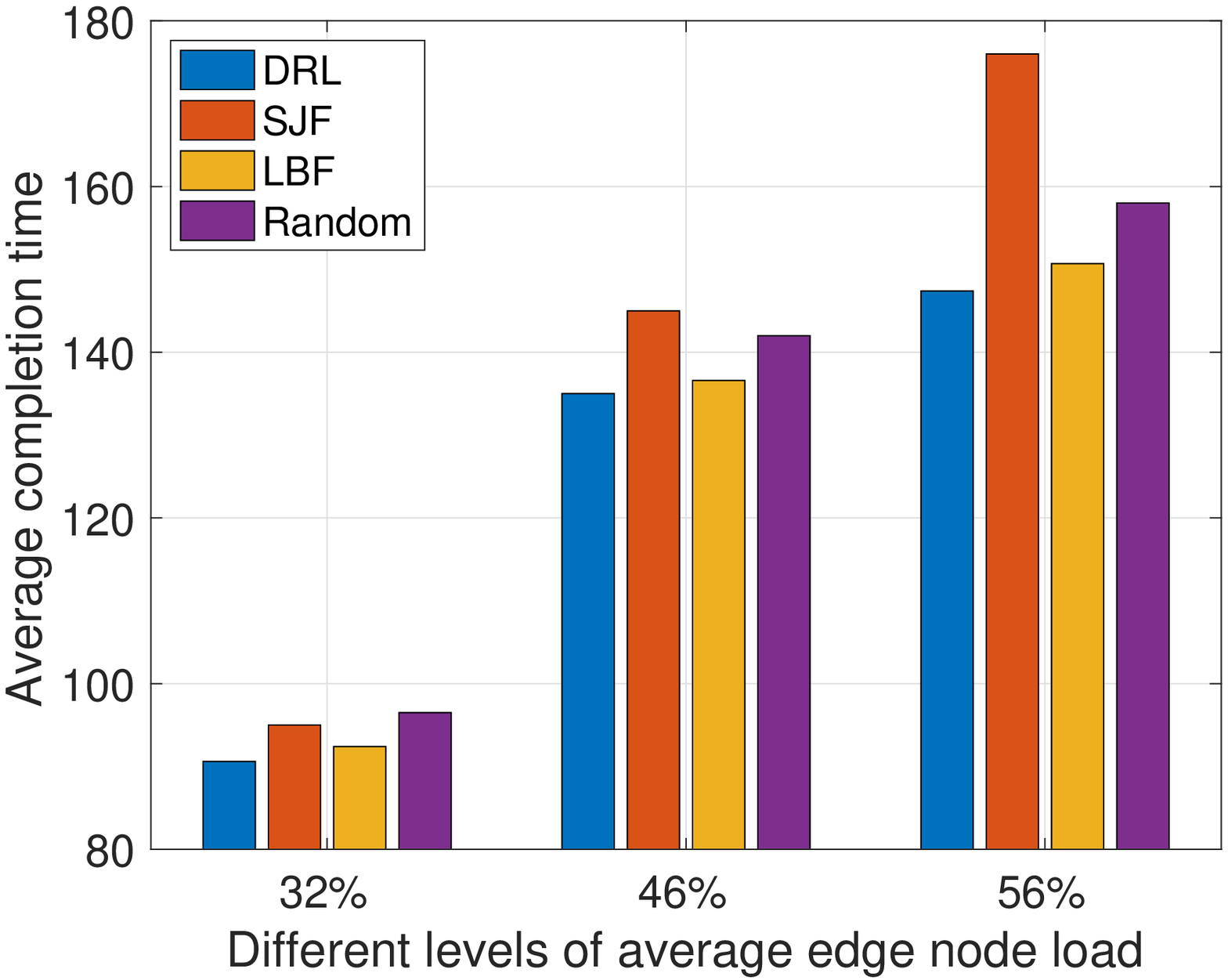}\\
				{}
			\end{center}
		\end{minipage}
	}
	\caption{Average slowdown $\bar{\phi}_g$ (left sub-figure) and completion time 
		$\bar{T}_g$ (right sub-figure) 
		at different levels of
		average edge node load, based on 
		dataset 1.
	}
	\label{dataset1}
\end{figure}

\begin{figure}[htb!]
	\centerline{
		\begin{minipage}{1.7in}
			\begin{center}
				\setlength{\epsfxsize}{1.7in}
				\epsffile{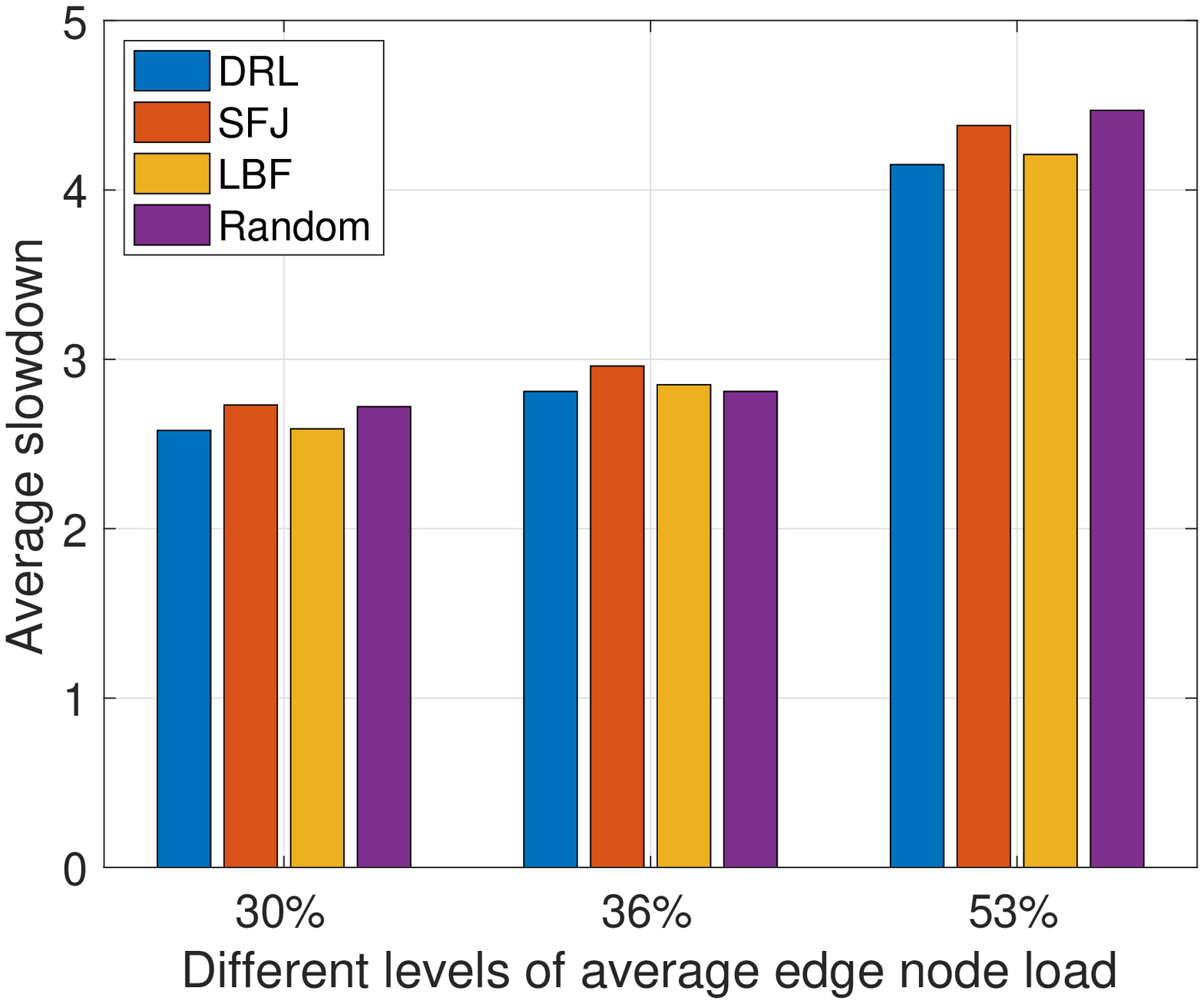}\\
				{}
			\end{center}
		\end{minipage}
		\begin{minipage}{1.7in}
			\begin{center}
				\setlength{\epsfxsize}{1.7in}
				\epsffile{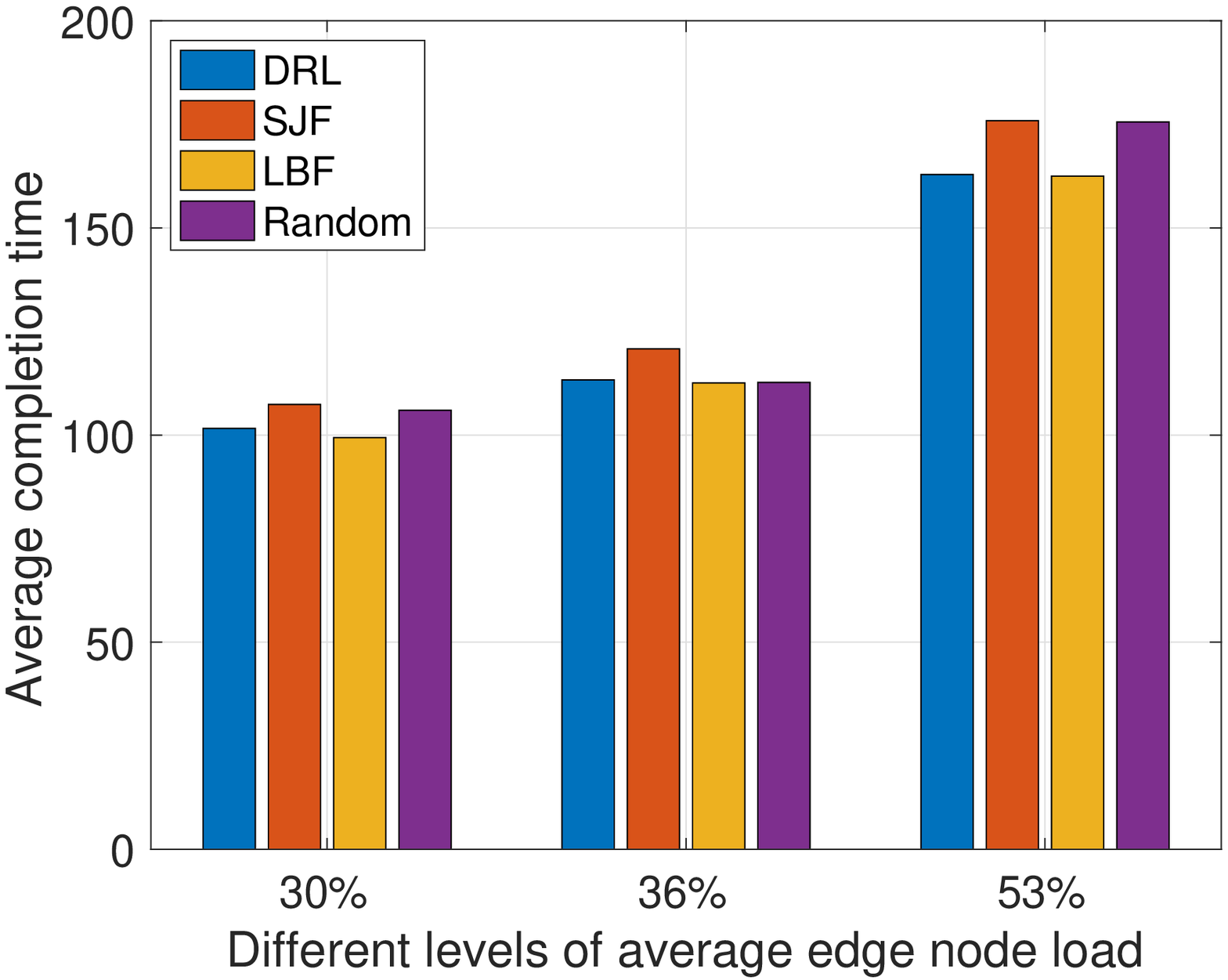}\\
				{}
			\end{center}
		\end{minipage}
	}
	\caption{Average slowdown $\bar{\phi}_g$ (left sub-figure) and completion time 
		$\bar{T}_g$ (right sub-figure) 
        at different levels of
		average edge node load, based on 
		dataset 2.
	}
	\label{dataset2}
\end{figure}

\begin{figure}[htb!]
	\centerline{
		\begin{minipage}{1.7in}
			\begin{center}
				\setlength{\epsfxsize}{1.7in}
				\epsffile{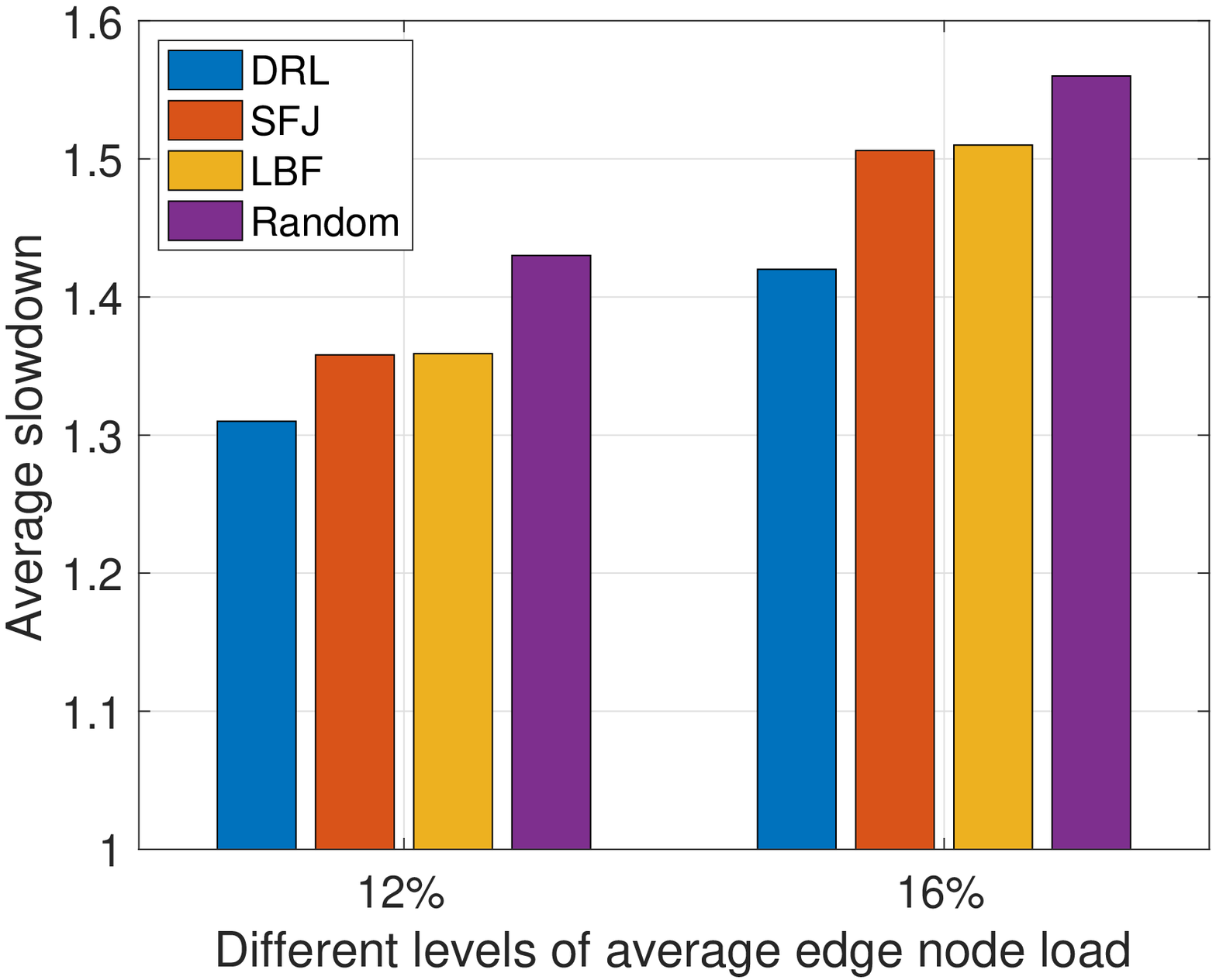}\\
				{}
			\end{center}
		\end{minipage}
		\begin{minipage}{1.7in}
			\begin{center}
				\setlength{\epsfxsize}{1.7in}
				\epsffile{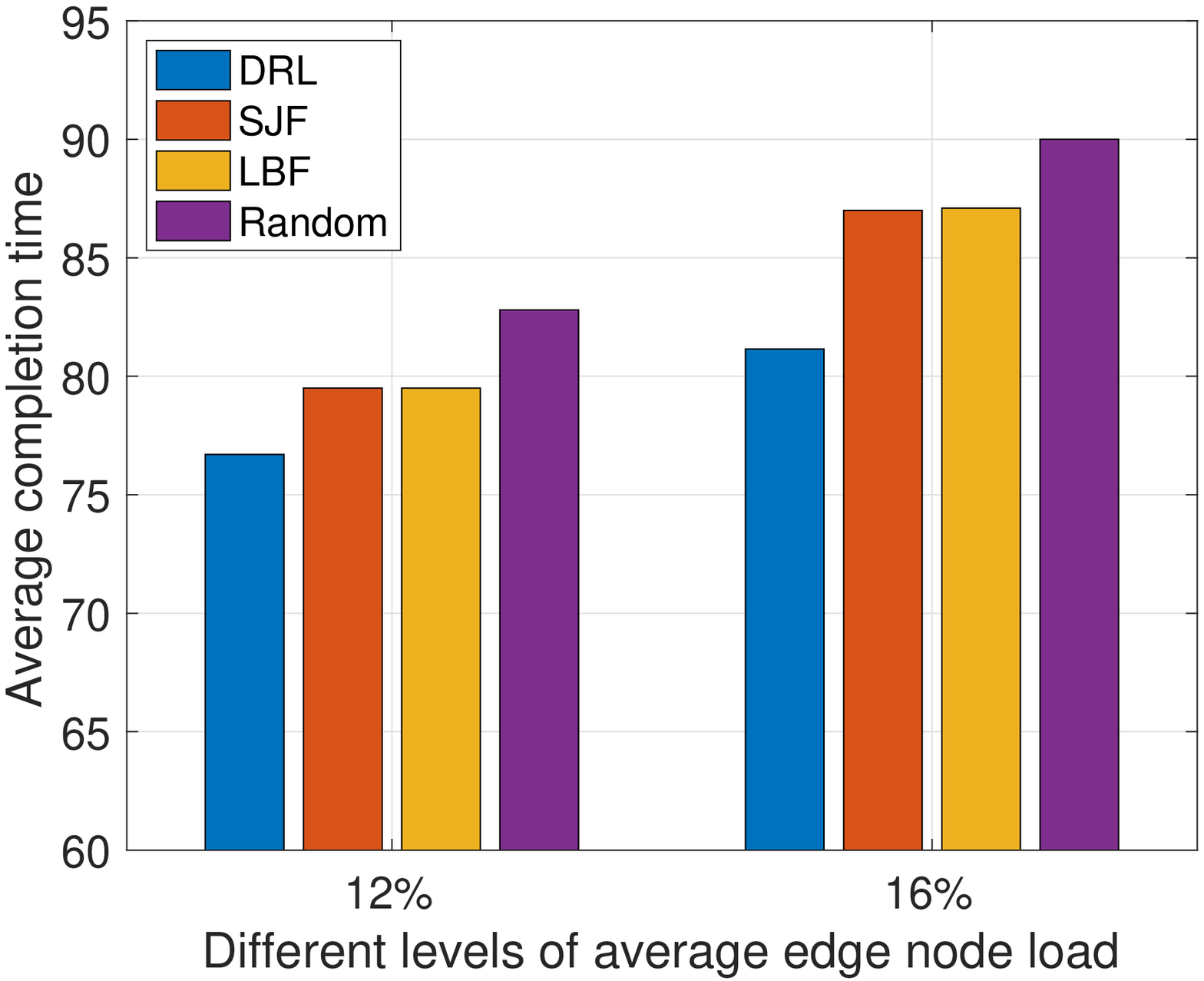}\\
				{}
			\end{center}
		\end{minipage}
	}
	\caption{Average slowdown $\bar{\phi}_g$ (left sub-figure) and completion time 
		$\bar{T}_g$ (right sub-figure) 
		at different levels of 
		average node load, based on 
		Google trace.
	}
	\label{dataset_google}
\end{figure}

Figures \ref{dataset1}, \ref{dataset2}, and \ref{dataset_google} 
illustrate the average slowdown $\bar{\phi}_g$ and the average completion time 
$\bar{T}_g$ of different designs based on three datasets and different average edge node load $\bar{L}$. 
We observe that as the workload increases, the average completion time and slowdown will also increase. 

In general, our DRL scheduling policy performs best among the four scheduling policies. 
For example, Figure \ref{dataset1} shows that 
the average slowdown of our DRL scheduling policy is $3.3\%$ less than LBF, $19.7\%$ 
less than SJF, and the average completion time of our DRL policy 
is $2\%$ less than LBF, and $19\%$ less than SJF when $\bar{L} = 56\% $. 
Regarding the simulations of Google data, Figure \ref{dataset_google} shows that our DRL scheduling policy 
has $5.6\%$ less average slowdown than SJF and LBF, and $7.4\%$ 
less completion time than SJF and LBF when $\bar{L} = 16\% $.
Additionally, simulations based upon Google data with $\bar{L}$ ranging from $28\%$ to $35\%$ are also performed. 
Our DRL scheduling policy shows clear advantage in average completion time and average slowdown compared with 
other three scheduling policies.

\section{Conclusion and Future Work}\label{sec_concl}

We introduced a Deep Reinforcement Learning model to address the scheduling of a collection 
of multi-component edge application jobs. Our DRL algorithm leverages the recent actor-critic method
to search for an online scheduling policy, which outperforms a few existing algorithms according to our 
experiments with synthetic data and a Google cloud data trace. Our preliminary study in this paper 
has shown the feasibility of applying the state-of-art DRL techniques to solve the complex resource 
allocation problems in edge computing systems. 

For future work, we will extend the model to include application component placement problem 
besides scheduling. 
Furthermore, we will extend our centralized DRL scheduling design 
to multi-agent collaborative DRL design.


\bibliographystyle{IEEEtran}
\bibliography{social_swarms}

\end{document}